\documentclass[12pt]{article}

\begin{document}

\begin{center}
{\bf Field theory of massive and massless vector particles in the
Duffin$-$Kemmer$-$Petiau formalism} \\
\vspace{5mm} S. I. Kruglov\\
%\footnote{E-mail: krouglov@utsc.utoronto.ca}
\vspace{3mm} \textit{University of Toronto at Scarborough,\\
Physical and Environmental Sciences Department, \\
1265 Military Trail, Toronto, Ontario, Canada M1C 1A4} \\

\vspace{5mm}
\end{center}

\begin{abstract}
Field theory of massive and massless vector particles is
considered in the first-order formalism. The Hamiltonian form of
equations is obtained after the exclusion of non-dynamical
components. We obtain the canonical and symmetrical Belinfante
energy-momentum tensors and their nonzero traces. We note that the
dilatation symmetry is broken in the massive case but in the
massless case the modified dilatation current is conserved. The
canonical quantization is performed and the propagator of the
massive fields is found in the Duffin$-$Kemmer$-$Petiau formalism.
\end{abstract}

\section{Introduction}

The theory of the unified weak and electromagnetic interaction
between elementary particles including vector particles (Standard
Model) is renormalized field theory \cite{Hooft}. The crucial role
plays the spontaneous breaking symmetry when massless vector
fields acquire masses due to the Higgs mechanism. Nowadays, it is
of great importance for searching the scalar Higgs bosons at Large
Hadronic Collider. Anyway, the old problem of describing massive
and massless vector particles is of theoretical interest.  It is
well known that the Proca equations for vector particles can be
cast into the first-order (the matrix form) Duffin$-$Kemmer$-$Petiau
(DKP) relativistic wave equation (RWE) \cite{Petiau},
\cite{Duffin}, \cite{Kemmer}, \cite{Umezawa}. One can
find early references on DKP equations in \cite{Krajcik}. The
matrix form of RWE is also convenient for the formulation of
higher derivative field equations \cite{Kruglov}, \cite{Kruglov(a)}, \cite{Kruglov(b)},
\cite{Kruglov(c)}, fields with
multi-spin \cite{Kruglov1}, \cite{Kruglov1(a)}, \cite{Kruglov1(b)},
\cite{Kruglov1(c)}, \cite{Kruglov1(d)}, \cite{Kruglov1'}, Einstein gravity
equations \cite{Fedorov1}, fields in curved space-time
\cite{Pimentel}, \cite{Pimentel1}, \cite{Bogush} and quantum chromodynamics
\cite{Gribov}. There is a vast number of papers devoted to DKP
equations, and, therefore, we mention only some part of them.

The goal of this paper is to give the systematic description of
massive and massless vector fields in the DKP formalism, to find
solutions in the form of projection matrix-dyads, to obtain the
quantum-mechanical Hamiltonian, to investigate the dilatation
symmetry, and to perform the canonical quantization.

The paper is organized as follows. In Sec.2, we consider massive
and massless vector fields in the form of RWE with two mass
parameters. Solutions of the wave equation for a free massive
field are obtained in the form of matrix-dyads. The
quantum-mechanical Hamiltonians are found for massive and massless
vector fields in Sec.3. We obtain the canonical and the
symmetrical Belinfante energy momentum tensors and the dilatation
current in Sec.4. It is demonstrated that the dilatation symmetry
is broken for massive fields but in the massless case the modified
dilatation current is conserved. In Sec.5, the canonical
quantization of massive fields is performed and the propagator of
fields is obtained in the DKP formalism. We draw a conclusion in
Sec.6. In Appendix, we demonstrate that the system of equations with
two mass parameter introduced can describe massive or massless fields.

The Euclidean metric is used and the system of units $\hbar =c=1$
is explored.

\section{Vector particles in the first-order formalism}

To consider the massive and massless vector fields, let us introduce the system of
equations with two mass parameters
\[
\partial _\nu \psi _{[\mu \nu ]}+m_1 \psi _\mu=0,
\]
\vspace{-8mm}
\begin{equation}
\label{1}
\end{equation}
\vspace{-8mm}
\[
\partial _\nu \psi _\mu -\partial _\mu \psi _\nu +m_2\psi _{[\mu \nu]}=0.
\]
System of equations (1) is the generalization of Proca equations, and describes massive or massless particles
depending on parameters $m_1$, $m_2$ chosen (see Appendix). At $m_1=m_2$, after the renormalization of fields, one arrives at the Proca equations. At $m_1=0$, we have the massless case corresponding to
the Maxwell equations. Excluding the antisymmetrical tensor $\psi
_{[\mu \nu ]}$ in Eq.(1) (see (A1) in Appendix), in the general case, we find the wave
equation for the field $\psi _\mu$ possessing the mass
$m=\sqrt{m_1m_2}$. Thus, Eq.(1) gives the convenient parametrization for description of
massive or massless vector fields. The fields $\psi_A$ ($A=\mu$,
$[\mu\nu]$) have the same dimension due to the presence of the
mass parameter $m_2$.

Introducing the wave function
\begin{equation}
\Psi (x)=\left\{ \psi _A(x)\right\} =\left(
\begin{array}{c}
\psi_\mu (x)\\
\psi_{[\mu\nu]}(x)\\
\end{array}
\right) \hspace{0.5in}(A=\mu , [\mu\nu]),\label{2}
\end{equation}
and the elements of the entire matrix algebra $\varepsilon
^{A,B}$, with properties
\begin{equation}
\left( \varepsilon ^{M,N}\right) _{AB}=\delta _{MA}\delta _{NB},
\hspace{0.5in}\varepsilon ^{M,A}\varepsilon ^{B,N}=\delta
_{AB}\varepsilon ^{M,N}, \label{3}
\end{equation}
where $A,B,M,N=\mu,[\mu\nu]$, the system of equations (1) can be
represented in the matrix form
\begin{equation}
\left[\partial _\mu \left(\varepsilon ^{\nu,[\nu\mu]}+ \varepsilon
^{[\nu\mu],\nu}\right) + m_1\varepsilon ^{\mu,\mu}+
m_2\frac{1}{2}\varepsilon ^{[\nu\mu],[\nu\mu]}\right] _{AB}\Psi
_B(x)=0 . \label{4}
\end{equation}
We imply a summation over all repeated indices. Defining the
$10\times 10$ matrices
\begin{equation}
\beta_\mu=\varepsilon ^{\nu,[\nu\mu]}+ \varepsilon
^{[\nu\mu],\nu},~~~\overline{P}=\varepsilon ^{\mu,\mu},~~~
P=\frac{1}{2}\varepsilon ^{[\nu\mu],[\nu\mu]}, \label{5}
\end{equation}
Eq.(4) takes the form of the first-order RWE
\begin{equation}
\left( \beta _\mu \partial _\mu + m_1\overline{P}+ m_2P\right)
\Psi (x)=0 . \label{6}
\end{equation}
The matrices $\beta_\mu$ are Hermitian matrices,
$\beta_\mu^+=\beta_\mu$. The projection operator $\overline{P}=
\overline{P}^+$ extracts the four-dimensional vector subspace
($\psi_\mu$) of the wave function $\Psi$, and the projection
operator $P=P^+$ extracts the six-dimensional tensor subspace
corresponding to the $\psi_{[\mu\nu]}$. The matrices $\beta _\mu$
obey the DKP algebra
\begin{equation}
\beta _\mu \beta _\nu \beta _\alpha +\beta _\alpha \beta _\nu
\beta _\mu =\delta _{\mu \nu }\beta _\alpha+\delta _{\alpha \nu
}\beta _\mu , \label{7}
\end{equation}
and matrices $\overline{P}$, $P$ are projection matrices
\[
\overline{P}^2=\overline{P},~~P^2=P,~~\overline{P}+P=1,~~
\overline{P}P=P\overline{P}=0,
\]
\vspace{-8mm}
\begin{equation}
\label{8}
\end{equation}
\vspace{-8mm}
\[
\beta_\mu\overline{P}+\overline{P}\beta_\mu=\beta_\mu,~~ \beta_\mu
P+P\beta_\mu=\beta_\mu.
\]
At $m_1=m_2=m$, from Eq.(6), we arrive at the DKP equation for
massive vector fields
\begin{equation}
\left( \beta _\mu \partial _\mu + m\right) \Psi (x)=0 . \label{9}
\end{equation}
For massless DKP equation, corresponding to the Maxwell equations,
we put $m_1=0$ in Eq.(6), and arrive at
\begin{equation}
\left( \beta _\mu \partial _\mu + m_2P\right) \Psi (x)=0 .
\label{10}
\end{equation}
In \cite{Harish}, \cite{Moroz} the case $m_1=0$, Eq.(10), was
considered. It should be noted that massless DKP model describes electromagnetic fields and is
invariant under a local U(1) gauge symmetry and being a fundamental
requirement for their description. The gauge transformations in DKP formalism are given by
\[
\Psi'=\Psi+\overline{P}\Phi.
\]
Then with the help of Eq.(8),(10), we obtain
\[
\left( \beta _\mu \partial _\mu + m_2P\right) \Psi'=(1-\overline{P})\beta _\mu \partial _\mu\Phi .
\]
Thus Eq.(10) is invariant under gauge transformations if the function $\Phi$ obeys the equation as follows (see also\cite{Harish}): $P\beta_\mu\partial_\mu\Phi=0$. It is easy to verify that this equation is valid for the function
\[
\Phi(x)=\left(
\begin{array}{c}
\partial_\mu \Lambda(x)\\
0
\end{array}\right),
\]
which leads to the gauge transformations: $\psi'_\mu(x)= \psi_\mu (x)+\partial_\mu \Lambda (x)$.
The gauge invariance of massless DKP equation was also discussed
in \cite{Lunardi}.

Now, we investigate the general case, Eq.(6), including two mass parameters, $m_1$ and $m_2$.
The Lorentz group generators in the $10$-dimension representation space are  given by
\begin{equation}
J_{\mu\nu}=\beta_\mu\beta_\nu-\beta_\nu\beta_\mu=
\varepsilon^{[\lambda\mu],[\lambda\nu]}+\varepsilon^{\mu,\nu}-
\varepsilon^{[\lambda\nu],[\lambda\mu]}-\varepsilon^{\nu,\mu},
 \label{11}
\end{equation}
and obey the commutation relations
\[
\left[J_{\rho\sigma},J_{\mu\nu}\right]=\delta_{\sigma\mu}J_{\rho\nu}+
\delta_{\rho\nu}J_{\sigma\mu}-\delta_{\rho\mu}J_{\sigma\nu}-\delta_{\sigma\nu}J_{\rho\mu},
\]
\vspace{-8mm}
\begin{equation}
\label{12}
\end{equation}
\vspace{-8mm}
\[
\left[\beta_{\lambda},J_{\mu\nu}\right]=\delta_{\lambda\mu}\beta_\nu
-\delta_{\lambda\nu}\beta_\mu,~~\left[\overline{P},J_{\mu\nu}\right]=
0,~~\left[P,J_{\mu\nu}\right]= 0 .
\]
Eq.(6) is form-invariant under the Lorentz transformations because
of Eq.(12). The Lorentz-invariant is $\overline{\Psi }\Psi =\Psi
^{+}\eta \Psi$, where $\Psi ^{+}$ is the Hermitian-conjugated wave
function, and the Hermitianizing matrix, $\eta $, is given by
\begin{equation}
\eta =\varepsilon ^{m,m}-\varepsilon ^{4,4} +\varepsilon
^{[m4],[m4]}-\frac{1}{2}\varepsilon ^{[mn],[mn]}. \label{13}
\end{equation}
The $\eta $ is the Hermitian matrix, $\eta^+ =\eta $, and obeys
the relations: $\eta \beta _m=-\beta _m\eta$ (m=1,2,3), $\eta
\beta _4=\beta _4\eta$. With the help of these relations, one
finds the "conjugated" equation
\begin{equation}
\overline{\Psi}(x)\left( \beta _\mu \overleftarrow{\partial}_\mu -
m_1\overline{P}- m_2P\right) =0 . \label{14}
\end{equation}
The Lagrangian leading to Eq.(6),(14) is
\[
\mathcal{L}=-\frac{1}{2}\overline{\Psi }(x)\left( \beta _\mu
\partial _\mu + m_1\overline{P} + m_2 P\right)\Psi (x)
\]
\vspace{-8mm}
\begin{equation}
\label{15}
\end{equation}
\vspace{-8mm}
\[
+ \frac{1}{2}\overline{\Psi }(x)\left(\beta _\mu
\overleftarrow{\partial _\mu} -m_1\overline{P}- m_2 P\right) \Psi
(x).
\]
In terms of fields $\psi_A$, the Lagrangian (15) reads
\begin{equation}
{\cal L}=\frac{1}{2}\left(\psi_{[\rho\mu]}^*\partial
_\mu\psi_\rho-\psi_\rho^*\partial _\mu\psi_{[\rho\mu]} - m_1
\psi_\mu^*\psi_\mu +
\frac{m_2}{2}\psi_{[\rho\mu]}^*\psi_{[\rho\mu]}\right)+ c.c.
,\label{16}
\end{equation}
where the $c.c.$ means the complex conjugated expression; complex
conjugation $*$ does not act on the metric imaginary unit $i$ of
fourth components in Eq.(16), $\psi_\mu^*=(\psi_m^*,i\psi_0^*)$,
and so on, and we used $\overline{\Psi
}=(\psi_\mu^*,-\psi_{[\mu\nu]}^*)$. It is easy to verify that
Euler-Lagrange equations $\partial{\cal
L}/\partial\psi_A-\partial_\mu\left(\partial{\cal L}/
\partial\partial_\mu\psi_A\right)=0$ lead to the equations of motion
 Eq.(1). The Lagrangian (15) (and (16)) vanishes for fields $\psi_A$ obeying Eq.(1) (or
Eq.(6),(14))).

\section{Solutions to the matrix equation}

Eq.(6), in the momentum space, for the positive ($+p$) and
negative ($-p$) energies, reads
\begin{equation}
\left( \pm i\widehat{p} + m_1\overline{P}+ m_2 P\right) \Psi (\pm
p)=0 , \label{17}
\end{equation}
where $\widehat{p}=\beta _\mu p _\mu$, and the four-momentum being
$p_\mu=(\textbf{p},ip_0)$ ($p^2= \textbf{p}^2-p_0^2$). One can
verify that the matrix of equation (17)
\begin{equation}
\Lambda_\pm = \pm i\widehat{p} + m_1\overline{P}+ m_2 P,
\label{18}
\end{equation}
obeys the ``minimal" matrix equation:
\begin{equation}
\left(\Lambda_\pm-m_1\right)\left(\Lambda_\pm-m_2 \right)
\left[\left(\Lambda_\pm-m_1\right)\left(\Lambda_\pm-m_2
\right)+p^2 \right]=0. \label{19}
\end{equation}
The non-trivial solutions to Eq.(17) exist if $\det \Lambda_\pm
=0$. Therefore, the eigenvalue of the matrix $\Lambda_\pm$ should
be zero, and this requirement results to the dispersion relation
\begin{equation}
p^2+m_1 m_2=0 . \label{20}
\end{equation}
Other eigenvalues of the matrix $\Lambda_\pm$ are $m_1$ and $m_2$,
that follows from Eq.(19). On-shell, $p^2=-m_1 m_2$, when $m_1\neq
m_2$, Eq.(19) becomes
\begin{equation}
\Lambda_\pm\left(\Lambda_\pm-m_1-m_2 \right)
\left(\Lambda_\pm-m_1\right)\left(\Lambda_\pm-m_2 \right)=0.
\label{21}
\end{equation}
Solutions to Eq.(17) in the form of the projection matrix
\cite{Fedorov} follow from Eq.(21):
\[
\Pi_\pm=N\left(\Lambda_\pm-m_1-m_2 \right)
\left(\Lambda_\pm-m_1\right)\left(\Lambda_\pm-m_2 \right)
\]
\vspace{-8mm}
\begin{equation}
\label{22}
\end{equation}
\vspace{-8mm}
\[
=\mp Ni\widehat{p}\left(\pm im_1\widehat{p}P\pm
im_2\widehat{p}\overline{P} -m_1 m_2\right),
\]
so that $\Lambda_\pm\Pi_\pm=0$, where $N$ is the normalization
constant. Every column of the matrix $\Pi_\pm$ is the solution to
Eq.(17). The projection matrix obeys the equation
\begin{equation}
\Pi_\pm^2=\Pi_\pm, \label{23}
\end{equation}
that leads to the normalization constant
\begin{equation}
N=-\frac{1}{m_1 m_2(m_1+m_2)} . \label{24}
\end{equation}
This can be verified with the help of Eq.(21). We notice that
Eq.(21) is the minimal polynomial only for the non-degenerate case
$m_1\neq m_2$. For the special case, $m_1=m_2$, we have to use the
minimal matrix equation on-shell:
\begin{equation}
\Lambda_\pm\left(\Lambda_\pm-m\right)\left(\Lambda_\pm-2m\right)=0.
\label{25}
\end{equation}
We obtain the projection operator from Eq.(25)
\begin{equation}
\Pi^{DKP}_\pm=\frac{1}{2m^2}\left(\Lambda_\pm^2-3m
\Lambda_\pm+2m^2\right) =\frac{\pm i\widehat{p}\left(\pm
i\widehat{p}-m\right)}{2m^2}. \label{26}
\end{equation}
Eq.(22) at $m_1=m_2$ becomes Eq.(26) .

We use the spin projection operators \cite{Fedorov}:
\begin{equation}
S_{(\pm 1)}=\frac 12\sigma _p\left( \sigma _p\pm 1\right)
,\hspace{0.5in}S_{(0)}=1-\sigma _p^2 .\label{27}
\end{equation}
where
\begin{equation}
\sigma _p=-\frac i{2\mid \mathbf{p}\mid }\epsilon
_{abc}p_aJ_{bc}=-\frac i{\mid \mathbf{p}\mid }\epsilon
_{abc}p_a\beta _b\beta _c. \label{28}
\end{equation}
One can check the relations $[\Lambda_\pm,\sigma _p]=0$, $S_{(\pm
1)}^2=S_{(\pm 1)}$, $S _{(\pm 1)}S_{(0)}=0$, $S_{(0)}^2=S_{(0)}$,
$[\Lambda_\pm,S_{(\pm 1)}]=0$, $[\Lambda_\pm,S_{(0)}]=0$. The
projection operators extracting solutions to Eq.(17) with spin
projections $\pm 1$, $0$ in the form of matrix-dyads are given by
\begin{equation}
\Pi_\pm S_{(\pm 1)}=\Psi_{\pm 1}\cdot \overline{\Psi}_{\pm 1}
,~~~~\Pi_\pm S_{(0)}=\Psi_{0}\cdot \overline{\Psi}_{0}. \label{29}
\end{equation}
The matrix-dyad has the matrix elements $(\Psi\cdot
\overline{\Psi})_{AB}= \Psi_A \overline{\Psi}_B $. Eq.(29) allow
us to make calculations of different electrodynamics processes
with vector particles in the covariant form \cite{Fedorov}.

For the case of massless vector particles (photons), the parameter
$m_1 =0$ and $p^2=0$. Then the matrix of equation (10) is
\begin{equation}
\Lambda^{(0)}_\pm= \pm i\widehat{p} + m_2 P, \label{30}
\end{equation}
and satisfies the minimal matrix equation
\begin{equation}
\Lambda^{(0)2}_\pm\left(\Lambda^{(0)2}_\pm-m_2\right)^2 =0.
\label{31}
\end{equation}
In this case zero eigenvalues of the operator $\Lambda^{(0)}_\pm $
are degenerated and it is impossible to obtain solutions in the
form of projection matrix-dyads \cite{Fedorov}, \cite{Moroz}. In
the case of generalized Maxwell equations such difficulty is
absent \cite{Kruglov2}, \cite{Kruglov2(a)}.

\section{The Hamiltonian form of equations}
\subsection{Massive fields}

It is very useful to obtain the quantum mechanical Hamiltonian
corresponding to equations (1) (or (6)) because the non-dynamical
components will be absent. There was a suggestion in \cite{Ghose}
to couple the electromagnetic field in the DKP equation only at
the level of the Hamiltonian form. Some aspects of Hamiltonian
form of DKP equations were considered in \cite{Nowakowski}, \cite{Nowakowski(a)},
\cite{Nowakowski(b)}. To
exclude the non-dynamical components, we rewrite Eq.(1) in the
form of two systems
\begin{equation}
m_2\psi _{[4 m]}=\partial _4 \psi _m -\partial _m \psi _4 , ~~~~
\partial _4 \psi _{[m 4 ]}+\partial _n \psi _{[m n]}+m_1 \psi _m =
0,
\label{32}
\end{equation}
\begin{equation}
m_2\psi _{[mn]}=\partial _m \psi _n -\partial _n \psi _m , ~~~~
\partial _n \psi _{[4n]}+m_1 \psi _4 = 0. \label{33}
\end{equation}
Equations (32) contain derivatives on the time of the dynamical
components $\psi_m$, $\psi_{[m4]}$, and Eq.(33) possess only
spatial derivatives on auxiliary (non-dynamical) components
$\psi_4$, $\psi_{[mn]}$. Replacing non-dynamical components from
Eq.(33) into Eq.(32), we arrive at the equations for the dynamical
components
\[
i\partial _t \psi _m=-m_2\psi _{[4 m]} +\frac{1}{m_1}\partial
_m\partial_n \psi _{[4 n]},
\]
\vspace{-8mm}
\begin{equation}
\label{34}
\end{equation}
\vspace{-8mm}
\[
i\partial _t \psi _{[m 4 ]}=\frac{1}{m_2}\left(\partial_m\partial
_n \psi _n-\partial_n^2\psi_m\right)+m_1 \psi _m .
\]
We can represent Eq.(34) in the matrix form introducing the
six-component wave function
\begin{equation}
\Phi (x)=\left(
\begin{array}{c}
\psi_n (x)\\
\psi_{[m 4]}(x)\\
\end{array}
\right).\label{35}
\end{equation}
With the help of the elements of the matrix algebra Eq.(3), we
rewrite Eq.(34) in the Schr\"{o}dinger-like form
\[
i\partial_t\Phi
(x)=\biggl[m_1\varepsilon^{[m4],m}-m_2\varepsilon^{n,[4n]}+\frac{1}{m_1}\varepsilon^{n,[4m]}\partial_n\partial_m
\]
\vspace{-7mm}
\begin{equation}
\label{36}
\end{equation}
\vspace{-7mm}
\[
+ \frac{1}{m_2}\biggl(\varepsilon^{[m4],n}\partial_m\partial_n\
-\varepsilon^{[m4],m}\partial_n^2\biggr)\biggr]\Phi (x),
\]
where the Hamiltonian is given by
\[
{\cal H}=m_1\varepsilon^{[m4],m}-m_2\varepsilon^{n,[4n]}
\]
\vspace{-8mm}
\begin{equation}
\label{37}
\end{equation}
\vspace{-8mm}
\[
+\frac{1}{m_1}\varepsilon^{n,[4m]}\partial_n\partial_m
+ \frac{1}{m_2}\biggl(\varepsilon^{[m4],n}\partial_m\partial_n\
-\varepsilon^{[m4],m}\partial_n^2\biggr).
\]
We have implied that $m_1\neq 0$, $m_2\neq 0$. The Hamiltonian
(37) is simplified for the choice $m_1=m_2\equiv m$:
\begin{equation}
{\cal H}=m\left(\varepsilon^{[m4],m}-\varepsilon^{n,[4n]}\right)+\frac{1}{m}\left[
\biggl(\varepsilon^{n,[4m]}+\varepsilon^{[m4],n}\biggr)\partial_m\partial_n\
-\varepsilon^{[m4],m}\partial_n^2\right]. \label{38}
\end{equation}
The six-component wave function (35) corresponds to three spin
states with positive (for particles) and negative (for
antiparticles) energies and does not contain auxiliary components.
We can introduce the minimal interaction of vector particles with
electromagnetic fields by the replacement $\partial_\mu\rightarrow
\partial_\mu-ieA_\mu$, where $A_\mu$ is the vector-potential of electromagnetic
fields. One may verify using the properties (3) that the matrix
Hamiltonian (37) in the momentum space obeys the minimal equation
\begin{equation}
\left({\cal H}^2- \textbf{p}^2-m_1m_2\right)\left({\cal H}^2+
\textbf{p}^2-m_1m_2\right)=0.\label{39}
\end{equation}
Equation (39) is an operator identity because in the momentum space, $\partial_\mu\rightarrow ip_\mu$,
the Hamiltonian is the matrix operator. If $m_1>0$ and $m_2>0$, the physical eigenvalue of the Hamiltonian
squared follows from Eq.(39): $p^2_0=\textbf{p}^2+m_1m_2$. The
projection operator extracting states with positive ($p_0$) and
negative ($-p_0$) energies is given by
\begin{equation}
\Sigma_{\pm}=\pm\frac{1}{4p_0\textbf{p}^2}\left({\cal H}\pm
p_0\right)\left({\cal H}^2+ \textbf{p}^2-m_1m_2\right),\label{40}
\end{equation}
so that $\Sigma_{\pm}^2=\Sigma_{\pm}$, $\Sigma_{\pm} {\cal
H}={\cal H}\Sigma_{\pm}=\pm p_0\Sigma_{\pm}$, where
$p_0=\sqrt{\textbf{p}^2+m_1m_2}$. The projection operator (40)
allows us to get solutions to the Hamiltonian equation (36).

\subsection{Massless fields}

In the case of massless vector fields (photons), we put $m_1=0$ in
Eq.(32),(33) and arrive at
\begin{equation}
m_2\psi _{[4 m]}=\partial _4 \psi _m -\partial _m \psi _4 , ~~~~
\partial _4 \psi _{[m 4 ]}+\partial _n \psi _{[m n]} =
0, \label{41}
\end{equation}
\begin{equation}
m_2\psi _{[mn]}=\partial _m \psi _n -\partial _n \psi _m , ~~~~
\partial _n \psi _{[4n]} = 0. \label{42}
\end{equation}
It is impossible to exclude the component $\psi_4$ from
Eq.(41),(42). Therefore, we need to consider the $\psi_4$ as a
dynamical component for the massless fields. To have the evolution
of the $\psi_4$ in time, we add to Eq.(41),(42) the Lorentz
condition $\partial_m\psi_m+\partial_4\psi_4=0$. After the
exclusion the non-dynamical component $\psi_{[mn]}$ from Eq.(41),
one finds the equations as follows:
\[
i\partial _t \psi _m=-m_2\psi _{[4 m]} -\partial_m \psi _4 ,
\]
\begin{equation}
i\partial _t \psi _4=\partial_n \psi _n , \label{43}
\end{equation}
\[
i\partial _t \psi _{[m 4 ]}=\frac{1}{m_2}\left(\partial_m\partial
_n \psi _n-\partial_n^2\psi_m\right) .
\]
Introducing the seven-component wave function
\begin{equation}
\Phi_0 (x)=\left(
\begin{array}{c}
\psi_\mu (x)\\
\psi_{[m 4]}(x)\\
\end{array}
\right),\label{44}
\end{equation}
and using the elements of the entire matrix algebra Eq.(3), we
represent Eq.(43) in the Hamiltonian form
\[
i\partial_t\Phi_0 (x)={\cal H}_0\Phi_0 (x),
\]
\vspace{-8mm}
\begin{equation}
\label{45}
\end{equation}
\vspace{-8mm}
\[
{\cal
H}_0=-m_2\varepsilon^{n,[4n]}+\left(\varepsilon^{4,m}-\varepsilon^{m,4}\right)\partial_m
+ \frac{1}{m_2}\biggl(\varepsilon^{[m4],n}\partial_m\partial_n\
-\varepsilon^{[m4],m}\partial_n^2\biggr).
\]
With the help of Eq.(3) one can check that the Hamiltonian (45) in
the momentum space satisfies the minimal equation
\begin{equation}
{\cal H}_0^2\left({\cal H}_0^2- \textbf{p}^2\right)=0.\label{46}
\end{equation}
The projection operator extracting states with positive
($p_0=|\textbf{p}|$) and negative ($p_0=-|\textbf{p}|$) energies
is
\begin{equation}
\Sigma^0_{\pm}=\pm\frac{1}{2|\textbf{p}|^3}\left({\cal H}_0\pm
|\textbf{p}|\right){\cal H}_0^2,\label{47}
\end{equation}
and $\left(\Sigma_{\pm}^0\right)^2=\Sigma^0_{\pm}$,
$\Sigma^0_{\pm} {\cal H}_0={\cal H}_0\Sigma^0_{\pm}=\pm|\textbf{p}|
\Sigma^0_{\pm}$, where $|\textbf{p}|=\sqrt{p_1^2+p_2^2+p_3^2}$.

\section{The energy-momentum tensor and dilatation current}

Now, we construct the energy-momentum tensor for a given Lagrangian (15).
With the help of the general equation \cite{Ahieser}
\begin{equation}
T^c_{\mu\nu}=\frac{\partial\mathcal{L}}{\partial\left(\partial_\mu
\Psi (x)\right)}\partial_\nu \Psi (x)+\partial_\nu \overline{\Psi
}(x) \frac{\partial\mathcal{L}}{\partial\left(\partial_\mu
\overline{\Psi} (x)\right)}-\delta_{\mu\nu} \mathcal{L}
,\label{48}
\end{equation}
we obtain the canonical energy-momentum tensor
\begin{equation}
T^c_{\mu\nu}=\frac{1}{2}\left(\partial_\nu \overline{\Psi}
(x)\right)\beta_\mu \Psi (x)-\frac{1}{2} \overline{\Psi}
(x)\beta_\mu \partial_\nu\Psi (x), \label{49}
\end{equation}
where we took into consideration that the Lagrangian (15) vanishes
for fields obeying the equations of motion. With the aid of
Eq.(2),(3),(5) the canonical energy-momentum tensor in the
component form becomes
\begin{equation}
T^c_{\mu\nu}=\frac{1}{2}\left(\psi_{[\rho\mu]}^\ast\partial_\nu
\psi_\rho-\psi_\rho^\ast\partial_\nu \psi_{[\rho\mu]}\right)+c.c.
. \label{50}
\end{equation}
It is easy to verify, using Eq.(1), that the energy-momentum
tensor (50) (and (49)) is conserved tensor, $\partial_\mu
T^c_{\mu\nu}=0$. The canonical energy-momentum tensor is not the
symmetric tensor, $T^c_{\mu\nu}\neq T^c_{\nu\mu}$, and its trace
is
\begin{equation}
T_{\mu\mu}^{c}=m_1\psi_\mu^\ast\psi_\mu-\frac{1}{2}m_2
\psi_{[\mu\nu]}^\ast\psi_{[\mu\nu]}. \label{51}
\end{equation}
One expects a nonzero trace of the energy-momentum
tensor and a broken dilatation current in a massive theory. But even for massless fields,
$m_1=0$, $m_2\neq 0$, the trace of the energy-momentum tensor in (51) is non-zero.
Therefore, it is of interest to investigate the scale invariance in the general case.
Thus, we consider the canonical dilatation current \cite{Coleman}
\begin{equation}
D_\mu^c=x_\alpha T_{\mu\alpha}^{c}+\Pi_\mu \textbf{d}\Psi,
\label{52}
\end{equation}
where
\begin{equation}
\Pi_\mu=\frac{{\partial\cal
L}}{\partial\left(\partial_\mu\Psi\right)} =-\overline{\Psi
}\beta_\mu, \label{53}
\end{equation}
and the matrix $\textbf{d}$ defines the field dimension. For the
Bose fields the $\textbf{d}$ is the unit matrix. From
Eq.(52),(53), we obtain
\begin{equation}
\partial_\mu D_\mu^c=T_{\mu\mu}^{c}, \label{54}
\end{equation}
where the conservation of the current was used
\begin{equation}
\partial_\mu j_\mu=\partial_\mu(i\overline{\Psi}\beta_\mu\Psi)=0. \label{55}
\end{equation}
The analogous relation follows from \cite{Coleman}
\[
\partial_\mu D_\mu^c=\Pi_\mu\left(\textbf{d}+1\right)\partial_\mu\Psi
 +\frac{\partial{\cal L}}{\partial\Psi}\textbf{d}\Psi
 +\overline{\Psi}\textbf{d}\frac{\partial{\cal L}}{\partial\overline{\Psi}}-4{\cal L}
\]
\vspace{-8mm}
\begin{equation}
\label{56}
\end{equation}
\vspace{-8mm}
\[
=\overline{\Psi
}\left(m_1\overline{P}+m_2P\right)\Psi=m_1\psi_\mu^\ast\psi_\mu-\frac{1}{2}m_2
\psi_{[\mu\nu]}^\ast\psi_{[\mu\nu]}.
\]
In Eq.(56), we took into account that for the charged particles
the $\Psi$ and $\overline{\Psi}$ are the independent wave
functions. The dilatation symmetry is broken because of massive
parameters $m_1$ and $m_2$. In the massless case, $m_1=0$, the
dilatation current $D_\mu^c$ is also not conserved, but later we
will introduce the modified conserved current.

To obtain the symmetrical the energy-momentum, we use the
expression for the Belinfante tensor \cite{Coleman}
 \begin{equation}
T_{\mu\alpha}^{B}=\frac{1}{2}\left(T_{\mu\alpha}^{c}+
\partial_\beta X_{\beta\mu\alpha}\right), \label{57}
\end{equation}
where
\begin{equation}
X_{\beta\mu\alpha}=\frac{1}{2}\left[\Pi_\beta J_{\mu\alpha}\Psi-
\Pi_\mu J_{\beta\alpha}\Psi-\Pi_\alpha
J_{\beta\mu}\Psi\right]+c.c.. \label{58}
\end{equation}
The additional complex conjugated term in Eq.(58) and the factor
$1/2$ in Eq.(57) are specific for our first-order formulation of
charged fields. From Eq.(2),(5),(53), we obtain the tensor
$X_{\beta\mu\alpha}$:
\begin{equation}
X_{\beta\mu\alpha}=\delta_{\alpha\beta}\psi_\lambda^\ast\psi_{[\lambda\mu]}-
\delta_{\alpha\mu}\psi_\lambda^\ast\psi_{[\lambda\beta]}
-\psi_\alpha^\ast\psi_{[\beta\mu]}+\psi_{[\mu\beta]}^\ast\psi_\alpha+c.c..
\label{59}
\end{equation}
With the help of expressions (59),(57), and equations of motion,
we obtain the Belinfante symmetric energy-momentum tensor
\[
T_{\mu\alpha}^{B}=m_2\left(\psi_{[\lambda\mu]}\psi_{[\alpha\lambda]}^\ast
+\psi_{[\lambda\mu]}^\ast\psi_{[\alpha\lambda]}\right)-m_1\left(\psi_\mu^
\ast\psi_\alpha +\psi_\alpha^\ast\psi_\mu\right)
\]
\vspace{-8mm}
\begin{equation}
\label{60}
\end{equation}
\vspace{-8mm}
\[
-\frac{1}{2}\delta_{\alpha\mu}\partial_\beta\left(\psi_\lambda^\ast\psi_{[\lambda\beta]}
+\psi_{[\lambda\beta]}^\ast\psi_\lambda\right).
\]
From Eq.(60), one finds the trace of the energy-momentum tensor
\begin{equation}
T_{\mu\mu}^{B}=2m_1\psi_\lambda^\ast\psi_\lambda .\label{61}
\end{equation}
In the case of massless fields, $m_1=0$, the trace of the
Belinfante symmetric energy-momentum tensor vanishes. We note that
the trace of the canonical energy-momentum tensor (51) does not
equal zero for massless fields. We evaluate a modified Belinfante
dilatation current \cite{Coleman}
 \begin{equation}
D_\mu^B=x_\alpha T_{\mu\alpha}^{B}+V_\mu, \label{62}
\end{equation}
where the field-virial $V_\mu$ is given by
 \begin{equation}
V_\mu=\Pi_\mu \Psi-\Pi_\alpha
J_{\alpha\mu}\Psi=-\overline{\Psi}\beta_\mu\Psi+\overline{\Psi
}\beta_\alpha J_{\alpha\mu}\Psi. \label{63}
\end{equation}
Using (2),(5),(11), one finds
\begin{equation}
\partial_\mu V_\mu=-m_1\psi_\lambda^\ast\psi_\lambda-\frac{1}{2}m_2
\psi_{[\mu\nu]}^\ast\psi_{[\mu\nu]}. \label{64}
\end{equation}
As a result of Eq.(61),(62),(64), we obtain
\begin{equation}
\partial_\mu D_\mu^B=T_{\mu\mu}^{B}+\partial_\mu V_\mu=m_1\psi_\lambda^\ast\psi_\lambda-\frac{1}{2}m_2
\psi_{[\mu\nu]}^\ast\psi_{[\mu\nu]}=\partial_\mu D_\mu^c.
\label{65}
\end{equation}
i.e. the same result (see Eq.(56)). For the massive fields,
$m_1\neq 0$, the dilatation symmetry is broken as it should be. For massless
fields ($m_1=0$), the currents $D_\mu^c$, $D_\mu^B$ also are not
conserved, but one can introduce the new conserved
current
\begin{equation}
D_\mu=D_\mu^B-V_\mu=x_\alpha T_{\mu\alpha}^{B}, \label{66}
\end{equation}
so that $\partial_\mu D_\mu=0$. Thus, massless fields (charged and
neutral electromagnetic fields) possess the dilatation symmetry
with the new dilatation current (66). Conformal symmetry of massless DKP equation also was
investigated in another formalism in \cite{Casana}.

\section{The canonical quantization of massive fields}

Let us consider the massive case implying that $m_1\neq 0$, $m_2\neq 0$.
The normalized solutions to Eq.(6) with definite energy-momentum
in the form of plane waves can be written as follows:
\begin{equation}
\Psi_s^{(\pm)}(x)=\sqrt{\frac{m_1+m_2}{2p_0 V}}\Psi_s(\pm
p)\exp(\pm ipx) , \label{67}
\end{equation}
where $p^2= \textbf{p}^2-p_0^2=-m_1m_2$, $V$ is the normalization
volume, and $s=\pm 1,0$ is the spin index. The function
$\Psi_s(\pm p)$ obeys Eq.(17), and we explore the normalization
conditions on the charge
\begin{equation}
\int_V \overline{\Psi}^{(\pm)}_{s}(x)\beta_4 \Psi^{(\pm)}_{s'
}(x)d^3 x=\pm\delta_{ss'} ,~~~~\int_V
\overline{\Psi}^{(\pm)}_{s}(x)\beta_4 \Psi^{(\mp)}_{s' }(x)d^3 x=0
, \label{68}
\end{equation}
where $\overline{\Psi}^{(\pm)}_{s}(x)=\left(\Psi^{(\pm)}_{s
}(x)\right)^+ \eta$. The field operators in the second quantized
theory are given by
\[
\Psi(x)=\sum_{p,s}\left[a_{p,s}\Psi^{(+)}_{s}(x) +
b^+_{p,s}\Psi^{(-)}_{s}(x)\right] ,
\]
\vspace{-7mm}
\begin{equation} \label{69}
\end{equation}
\vspace{-7mm}
\[
\overline{\Psi}(x)=\sum_{p,s}\left[a^+_{p,s}
\overline{\Psi}^{(+)}_{s}(x)+ b_{p,s}
\overline{\Psi}^{(-)}_{s}(x)\right] ,
\]
where the positive and negative parts of the wave function
$\Psi_s^{(\pm)}(x)$ are defined by Eq.(67). The creation and
annihilation operators of particles, $a^+_{p,s}$, $a_{p,s}$ and
antiparticles $b^+_{p,s}$, $b_{p,s}$ obey the usual commutation
relations
\[
[a_{p,s},a^+_{p',s'}]=\delta_{ss'} \delta_{pp'} ,~~~
[a_{p,s},a_{p',s'}]=[a^+_{p,s},a^+_{p',s'}]=0,
\]
\begin{equation}
[b_{p,s},b^+_{p',s'}]=\delta_{ss'} \delta_{pp'} ,~~~
[b_{p,s},b_{p',s'}]=[b^+_{p,s},b^+_{p',s'}]=0, \label{70}
\end{equation}
\[
[a_{p,s},b_{p',s'}]=[a_{p,s},b^+_{p',s'}]=
[a^+_{p,s},b_{p',s'}]=[a^+_{p,s},b^+_{p',s'}]=0.
\]
It follows from commutation relations (70) that the metric is positive-definite.
Thus, we construct vector space which is really a Hilbert space.
For quantization of the massless fields, one needs to introduce indefinite metric (see Ref. \cite{Kruglov2(a)}).
From Eq.(68)-(70), one can find the commutation relations for different times
\[
\left[\Psi_M(x),\Psi_N(x')]=[\overline{\Psi}_{ M}(x),
\overline{\Psi}_{ N}(x')\right] =0, ~~\left[\Psi_{
M}(x),\overline{\Psi}_{N}(x')\right]=N_{ MN}(x,x'),
\]
\[
N_{MN}(x,x')=N^+_{MN}(x,x')-N^-_{ MN}(x,x'),
\]
\vspace{-7mm}
\begin{equation} \label{71}
\end{equation}
\vspace{-7mm}
\[
N^+_{MN}(x,x')=\sum_{p,s}\left(\Psi^{(+)}_{s}(x)\right)_M
\left(\overline{\Psi}^{(+)}_{s}(x')\right)_N ,
\]
\[
N^-_{MN}(x,x')=\sum_{p,s}\left(\Psi^{(-)}_{s}(x)\right)_M
\left(\overline{\Psi}^{(-)}_{s}(x')\right)_N .
\]
We obtain from Eq.(67),(71)
\begin{equation}
N^\pm_{MN}(x,x')=\sum_{p,s}\frac{m_1+m_2}{2p_0
V}\left(\Psi_{s}(\pm p)\right)_M\left(\overline{\Psi}_{s}(\pm
p)\right)_N\exp [\pm ip(x-x')] .
 \label{72}
\end{equation}
From Eq.(27), one finds that the equation
\begin{equation}
S_{(+ 1)}+ S_{(- 1)}+S_{(0)}=1,\label{73}
\end{equation}
is valid. Then, from Eq.(22),(29), we obtain
\begin{equation}
\Pi_{\pm}=\sum_s\left(\Psi_{s}(\pm
p)\right)\cdot\left(\overline{\Psi}_{s}(\pm p)\right) = \frac{\pm
i\widehat{p}\left(\pm im_1\widehat{p}P\pm
im_2\widehat{p}\overline{P}
-m_1m_2\right)}{m_1m_2\left(m_1+m_2\right)}. \label{74}
\end{equation}
and Eq.(72) reads
\[
N^\pm_{MN}(x,x')=\sum_{p} \frac{1}{2p_0 V} \left(\frac{\pm
i\widehat{p}\left(\pm im_1\widehat{p}P\pm
im_2\widehat{p}\overline{P} -m_1 m_2\right)}{m_1 m_2}\right)_{MN}
\exp [\pm ip(x-x')]
\]
\vspace{-6mm}
\begin{equation} \label{75}
\end{equation}
\vspace{-6mm}
\[
= \left( \frac{\beta_\mu\partial_\mu\left(m_1\beta_\mu
P\partial_\mu + m_1\beta_\mu \overline{P}\partial_\mu -m_1
m_2\right)}{m_1 m_2} \right)_{MN} \Delta_{\pm}(x-x'),
\]
where we introduce the singular functions \cite{Ahieser}
\begin{equation}
\Delta_+(x)=\sum_{p}\frac{1}{2p_0V}\exp
(ipx),~~~~\Delta_-(x)=\sum_{p}\frac{1}{2p_0V}\exp (-ipx).
\label{76}
\end{equation}
With the aid of the function \cite{Ahieser}
\begin{equation}
\Delta_0 (x)=i\left(\Delta_+(x)-\Delta_-(x)\right), \label{77}
\end{equation}
and using Eq.(71),(75)-(77), one finds
\begin{equation}
N_{MN}(x,x')=-i\left(\frac{\beta_\mu\partial_\mu\left(m_1\beta_\mu
P\partial_\mu + m_2\beta_\mu \overline{P}\partial_\mu -m_1
m_2\right)}{m_1 m_2}\right)_{MN} \Delta_{0}(x-x').
 \label{78}
\end{equation}
The function $\Delta_{0}(x)$ vanishes when $x^2=
\textbf{x}^2-t^2>0$ \cite{Ahieser}. One can check with the help of
Eq.(7), that the relation
\[
 \left(\beta_\mu\partial_\mu+ m_1
\overline{P}+m_2 P
\right)\frac{\beta_\mu\partial_\mu\left(m_1\beta_\mu P\partial_\mu
+ m_2\beta_\mu \overline{P}\partial_\mu -m_1 m_2\right)}{m_1 m_2}
\]
\vspace{-7mm}
\begin{equation} \label{79}
\end{equation}
\vspace{-7mm}
\[
= \frac{\left( m_1 \overline{P}+m_2 P\right)\beta_\mu\partial_\mu
}{m_1 m_2}\left(\partial_\mu^2-m_1 m_2\right).
 \]
holds. Using the equation \cite{Ahieser} $\left(\partial_\mu^2-m_1
m_2\right)\Delta_\pm (x)=0$, from Eq.(75),(79), we arrive at
\begin{equation}
\left(\beta_\mu\partial_\mu + m_1 \overline{P}+m_2
P\right)N^\pm(x,x')=0 .
 \label{80}
\end{equation}
The vacuum expectation of chronological pairing of operators (the
propagator) is defined as \cite{Ahieser}
\[
\langle T\Psi_{M}(x)\overline{\Psi}_{ N}(y)\rangle_0=N^c_{
MN}(x-y)
\]
\vspace{-8mm}
\begin{equation} \label{81}
\end{equation}
\vspace{-8mm}
\[
=\theta\left(x_0 -y_0\right)N^+_{MN}(x-y)+\theta\left(y_0
-x_0\right)N^-_{MN}(x-y) ,
\]
where $\theta(x)$ is the theta-function. Then, one obtains from
Eq.(81) the propagator:
\begin{equation}
\langle T\Psi(x)\cdot\overline{\Psi}(y)\rangle_0
=\frac{\beta_\mu\partial_\mu\left(m_1\beta_\mu P\partial_\mu +
m_2\beta_\mu \overline{P}\partial_\mu -m_1 m_2\right)}{m_1
m_2}\Delta_c (x-y),
 \label{82}
\end{equation}
where
\begin{equation}
\Delta_c (x-y)=\theta\left(x_0
-y_0\right)\Delta_+(x-y)+\theta\left(y_0 -x_0\right)\Delta_-(x-y)
. \label{83}
\end{equation}
With the help of the equation \cite{Ahieser}
$\left(\partial_\mu^2-m_1 m_2\right)\Delta_c (x)=i\delta(x)$, and
from Eq.(79),(82),(83), we find
\begin{equation}
\left(\beta_\mu\partial_\mu+ m_1 \overline{P}+m_2 P \right)\langle
T\Psi(x)\cdot\overline{\Psi}(y)\rangle_0 = i\frac{ m_1
\overline{P}+m_2 P }{m_1 m_2}\beta_\mu\partial_\mu\delta (x-y).
 \label{84}
\end{equation}
At the chose $m_1=m_2=m$, and taking into consideration Eq.(8),
$P+\overline{P}=1$, the propagator (82) and Eq.(84) are
simplified. The above equations are not valid for the massless
case, $m_1=0$, because of singularities in Eq.(78),(79),(82),84). For the massive case, the
propagator (82) can be used for calculations of quantum processes
with vector particles in the first-order formalism.
The difficulty to quantize massless fields in the 10-component form of equations is connected with
the fact that eigenvalues of the operator $\Lambda^{(0)}_\pm $, Eq.(31), are
degenerated. To avoid this difficulty, one needs to introduce a general gauge and
to use the 11-component RWE. Quantization of massless fields (in
the general gage) in the first-order formalism was performed in Ref. \cite{Kruglov2(a)}.

\section{Conclusion}

We have considered the massive and massless vector fields in the
DKP formalism. Solutions in the form of matrix-dyads obtained
allow us to make quantum-electrodynamics calculations of processes
with vector particles in the covariant form. After the exclusion
of the non-dynamical components the Hamiltonian forms of equations
for massive and massless fields are obtained. One may consider
particles in external electromagnetic fields at the level of the
Hamiltonian form. The canonical and symmetrical Belinfante
energy-momentum tensors found possess their nonzero traces. We
investigate the dilatation symmetry in the first-order formalism.
It was demonstrated that the dilatation symmetry is broken in the
massive case but in the massless case the modified dilatation
current is conserved. The canonical quantization is considered in
the DKP form and the propagator of the massive fields is obtained.
Quantization of fields and the propagator found make it possible
to use the perturbation theory for different quantum calculations
in a simple manner.

\vspace{5mm}

\textsc{\textbf{Appendix}}

\vspace{5mm}

Now, we clarify the physical meaning of parameters $m_1$, $m_2$ introduced in Eq.(1).
Eliminating the antisymmetric tensor in Eq.(1), one obtains an equation as follows:
\[
\left(\delta_{\mu\nu}\partial_\alpha^2-\partial_\mu\partial_\nu\right)\psi_\nu-m_1m_2\psi_\mu=0,~~~~~~~~~~~~~~~
~~~~~~~~~~~~~~~~~~~~~~~~~~~~~~~~~~(A1)
\]
Let us consider the momentum space, $\partial_\mu\rightarrow ip_\mu$. Then Eq.(A1)
becomes the matrix equation
\[
M_{\mu\nu}\psi_\nu=0,~~~~~~~~~~~~~~~~~~~~~~~~~~~~~~~~~~~~~~~~~~~~~~~~~~~~~~~~~~~~~~~~~~~~~~~~~~~~~~~~~~~~(A2)
\]
where the matrix of the equation is given by
\[
M_{\mu\nu}=\delta_{\mu\nu}\left(p^2+m_1m_2\right)-p_\mu p_\nu,~~~~~~~~~~~~~~~~~~~~~~~~~~~~~~~~~~~~~~~~~~~~~~~~~~~(A3)
\]
where $p^2=p_\mu^2=\textbf{p}^2-p_0^2$. One can verify that the matrix $M=\{M_{\mu\nu}\}$ obeys the ``minimal" polynomial equation
\[
\left(M-p^2-m_1m_2\right)\left(M-m_1m_2\right)=0.~~~~~~~~~~~~~~~~~~~~~~~~~~~~~~~~~~~~~~~~~~~~~~~~~~~~~~~~~~~~~~(A4)
\]
It follows from Eq.(A4) that the matrix $M$ has the eigenvalues $\lambda_1=p^2+m_1m_2$ and $\lambda_1=m_1m_2$.
The algebraic matrix equation (A2) possesses non-trivial solutions if $\det M=0$. This leads to $\lambda_1=0$ that is equivalent to the dispersion equation
\[
p_0^2=\textbf{p}^2+m_1m_2,~~~~~~~~~~~~~~~~~~~~~~~~~~~~~~~~~~~~~~~~~~~~~~~~~~~~~~~~~~~~~~~~~~~~~~~~~~~~~~~~~~~~~(A5)
\]
or $\lambda_2=0$. Eq.(A5) tells us that the mass of a particle is $m=\sqrt{m_1m_2}$. The equation $\lambda_2=0$ is satisfied if $m_1=0$ (we do not discuss the trivial case $m_2=0$). Equation $m_1=0$ corresponds to massless case because it leads to $p_0^2=\textbf{p}^2$. It should be noted that Eq.(A1) (or system (1)) does not describe simultaneously two particles (massive and massless). Eq.(A1) describes massive particles if $m_1\neq 0$, $m_2\neq 0$ or massless particles if $m_1= 0$, $m_2\neq 0$. Thus, we have the convenient parametrization which allows us to consider both cases: massive and massless. Applying the 4-divergence to Eq.(A1), one finds that the equation $\partial_\mu\psi_\mu=0$ holds automatically only for massive case $m_1\neq 0$, $m_2\neq 0$. The equation $\partial_\mu\psi_\mu=0$ leaves only three physical degrees from 4-vector $\psi_\mu$ corresponding to three spin projections of vector particles. For massless fields ($m_1=0$) the Lorentz condition $\partial_\mu\psi_\mu=0$ does not follow from the equation of motion (A1) which becomes the wave equation for the four-potential \cite{Itzykson}.


\begin{thebibliography}{99}

\bibitem{Hooft} Gerard 't Hooft, M. J. G. Veltman,
Nucl. Phys. \textbf{B44}, 189 (1972).

\bibitem{Petiau} G. Petiau, University of Paris thesis (1936).
% Acad. Roy. de Belg., Classe Sci., Mem in 8o 16, No. 2 (1936).

\bibitem{Duffin} R. J. Duffin, Phys. Rev. \textbf{54}, 1114 (1938).

\bibitem{Kemmer} N. Kemmer, Proc. Roy. Soc. \textbf{A173}, 91 (1939).

 \bibitem{Umezawa} H. Umezawa, Quantum Field Theory (North-Holland, 1956).

\bibitem{Krajcik} R. A. Krajcik and M. M. Nieto, Am. J. Phys. \textbf{45}, 818 (1977).

\bibitem{Kruglov} S. I. Kruglov, Annales Fond. Broglie \textbf{29}, 1005 (2004)
[Erratum-ibid \textbf{31}, 489 (2006)] [arXiv:quant-ph/0408056].
\bibitem{Kruglov(a)} S. I. Kruglov, Annales Fond. Broglie \textbf{31}, 343,(2006)
[arXiv:hep-th/0606128].
\bibitem{Kruglov(b)} S. I. Kruglov, Can. J. Phys. \textbf{85}, 887 (2007)
[arXiv:hep-ph/0507027].
\bibitem{Kruglov(c)} S. I. Kruglov, J. Phys. \textbf{A43}, 245403 (2010)
[arXiv:0907.1706 [hep-th]].

\bibitem{Kruglov1} S. I. Kruglov, Int. J. Mod. Phys. \textbf{A16}, 4925 (2001)
[arXiv:hep-th/0110083].
\bibitem{Kruglov1(a)} S. I. Kruglov, Int. J. Mod. Phys. \textbf{A21}, 1143 (2006)
[arXiv:hep-th/0405088].
\bibitem{Kruglov1(b)} S. I. Kruglov, Mod. Phys. Lett. \textbf{A23}, 2141 (2008)
[arXiv:0802.0256 [hep-th]].
 \bibitem{Kruglov1(c)} S. I. Kruglov, Int. J. Theor. Phys. \textbf{41}, 653, (2002)
[arXiv:hep-th/0110251].
\bibitem{Kruglov1(d)} S. I. Kruglov, Eur. Phys. J. \textbf{C68}, 337 (2010)
[arXiv:0911.0442 [hep-th]].

\bibitem{Kruglov1'} S. I. Kruglov, Symmetry and Electromagnetic Interaction of Fields with
Multi-Spin (Nova Science Publishers, Huntington, New York,
(2001)).

\bibitem{Fedorov1} F. I. Fedorov, Doklady Acad. Sci. USSR \textbf{179}, 802
(1968).

\bibitem{Pimentel} R. Casana, J. T. Lunardi, B. M. Pimentel, R. G. Teixeira,
Gen. Rel. Grav. \textbf{34}, 1941 (2002)
[arXiv:gr-qc/0203068].
\bibitem{Pimentel1} R. Casana, V. Y. Fainberg, J. T. Lunardi, B. M. Pimentel, R. G. Teixeira,
Class. Quant. Grav. \textbf{20}, 2457 (2003)
[arXiv:gr-qc/0209083].

\bibitem{Bogush} A. A. Bogush, V. V. Kisel, N. G. Tokarevskaya, V. M. Red'kov,
Annales Fond. Broglie \textbf{32}, 355 (2007).

\bibitem{Gribov} V. Gribov, Eur. Phys. J. \textbf{C10}, 71 (1999)
[arXiv:hep-ph/9807224].

%\bibitem{Bogush1} A. A. Bogush and L. G. Moroz. Vvedenie v teoriu klassicheskih
%polei (Nauka i Tekhnika, Minsk, 1968) (in Russian).

\bibitem{Harish} Harish-Chandra, Proc. Roy. Soc. \textbf{A186}, 502 (1946).

\bibitem{Moroz} L. G. Moroz and F. I. Fedorov, Proc. of Institute
of Physics and Mathematics, Acad. Nauk BSSR, \textbf{3}, 154 (1959) (in Russian).

\bibitem{Lunardi} J. T. Lunardi, B. M. Pimentel, R. G. Teixeira, J. S. Valverde,
Phys. Lett. \textbf{A268}, 165 (2000)
[arXiv:hep-th/9911254].

\bibitem{Fedorov} F. I. Fedorov, Sov. Phys. - JETP \textbf{35}(8), 339 (1959) (Zh.
Eksp. Teor. Fiz. \textbf{35}, 493 (1958)).

\bibitem{Kruglov2} S. I. Kruglov, Annales Fond. Broglie {\bf 26}, 725 (2001)
[arXiv:math-ph/0110008].
\bibitem{Kruglov2(a)} S. I. Kruglov, Can. J. Phys. \textbf{86}, 995 (2008)
[Erratum-ibid. {\bf 89}, 249 (2011)]
[arXiv:hep-th/0610202].

\bibitem{Ghose} P. Ghose, Foundations of Physics \textbf{26}, 1441 (1996).

\bibitem{Nowakowski} L. Oliver, Anales de Fisica, \textbf{64}, 407 (1968).
\bibitem{Nowakowski(a)} M. Nowakowski, Phys. Lett. \textbf{A244}, 329 (1998)
[arXiv:quant-ph/9801020].
\bibitem{Nowakowski(b)} T. R. Cardoso, L. B. Castro and A. S. de Castro, Phys. Lett. \textbf{A 372},
5964 (2008)
[arXiv:0806.0359 [hep-th]].

%\bibitem{Kanatchikov} I. V. Kanatchikov, Rep. Math. Phys. \textbf{46}, 107 (2000) (hep-th/9911175).

\bibitem{Ahieser} A. I. Ahieser and V. B. Berestetskii, Quantum
Electrodynamics (New York: Wiley Interscience, 1969).

\bibitem{Coleman} S. Coleman and R. Jakiw, Ann. Phys. \textbf{67}, 552 (1971).

\bibitem{Casana} R. Casana, B. M. Pimentel, J. T. Lunardi, R. G. Teixeira,
Class. Quantum Grav. \textbf{22}, 3083 (2005)
[arXiv:gr-qc/0311042].

\bibitem{Itzykson} C. Itzykson and J. B. Zuber, Quantum Field Theory (McGraw-Hill Inc., 1980).

\end{thebibliography}
\end{document}